\documentclass[runningheads]{llncs}

\usepackage{graphicx}
\usepackage{amsmath,amssymb,amstext,times}
\usepackage{color}
\usepackage{float}
\usepackage{booktabs}
\usepackage{hyperref}
\usepackage{multirow}
\usepackage{comment}
\usepackage{algorithm,algorithmic}
\makeatletter
\newcommand{\ALOOP}[1]{\ALC@it\algorithmicloop\ #1%
  \begin{ALC@loop}}
\newcommand{\ENDALOOP}{\end{ALC@loop}\ALC@it\algorithmicendloop}

\makeatother

\newcommand{\bg}[1]{\boldsymbol{#1}}
\newcommand{\bm}[1]{\mathbf{#1}} 

\newcommand\raiseT[2]{%
\setbox0\hbox{$#1{#2}$}\raise\dp0\box0}

\usepackage{todonotes}

\begin{document}
\title{WaveletBrain: Characterization of human brain via spectral graph wavelets}
%
%
\author{
 Majid Masoumi \and  
 Matthew Toews \and  
 Herv\'e Lombaert,
 for the Alzheimer's Disease Neuroimaging Initiative \thanks{Corresponding Author: ma\_masou@encs.concordia.ca. 
Data used in preparation of this article were obtained from the Alzheimer's Disease Neuroimaging Initiative (ADNI) database (adni.loni.usc.edu). As such, the investigators within the ADNI contributed to the design and implementation of ADNI and/or provided data but did not participate in analysis or writing of this report. A complete listing of ADNI investigators can be found at: \url{http://adni.loni.usc.edu/wp-content/uploads/how_to_apply/ADNI_Acknowledgement_List.pdf}}}
 
 \authorrunning{M. Masoumi et al.}
 \institute{\'Ecole de technologie sup\'erieure, Montreal, Canada \\}
%
\maketitle              
\begin{abstract}
Early diagnosis of Alzheimer's disease plays a key role in understanding the degree of the patient's mental decline and determining preventive therapies. In this study, we introduce WaveletBrain, a novel representation of the white and gray matter surfaces of the cortex. The proposed framework innovates by deriving localized shape information from a global harmonic representation, that can be used in large-scale population studies of surface data. Results show that WaveletBrain leads to statistically significant improvements in comparison to the ShapeDNA representation in a variety of experiments including (i) classification of Alzheimer's disease, normal aging, and mild cognitive impairment, (ii) sex classification and (iii) age prediction of subjects. We performed our analysis on 719 patients and 2,876 surfaces. While this work focuses primarily on Alzheimer's disease diagnosis, our proposed framework can be used to address general surface analysis problems in neuroscience.

\end{abstract}

\section{Introduction}
 Alzheimer's disease (AD) is a progressive disease that affects the brain both mentally and physically. In fact, the brain morphometry starts to change years before recognizing the first symptoms of AD. Although no cure for AD is known, efforts towards early detection of the disease could directly impact the quality of a patient's life \cite{Paquerault:12}.

Recently, there has been a flurry of research activity on designing a computer-assisted model to detect subtle changes in brain shape and, hence determining risk level and preventive therapiess \cite{Falahati:14,wachinger:16,wachinger:15}. Measuring the similarity between all image pairs of a large set of N magnetic resonance images (MRIs) is a computationally expensive operation of $O(N^2)$ complexity \cite{toews:15}. Moreover, scanning artifacts and inhomogeneities can lead to noisy intensity variations in MRI scans, limiting the accuracy of AD classification \cite{wachinger:15}. The accuracy of prediction is closely linked to the model or representation used \cite{Dickinson:14}. The cortical surface is responsible for all cognition, and finding an efficient and informative representation may lead to significant improvements in tasks such as classification and regression. Early approaches to modeling brain shape included volumetric measurements or voxel-wise morphometric analysis \cite{Good:01,Shi:09,Hastings:04}. Although these representations are useful in describing inter-class differences, they ignore important regional information about brain shape \cite{shakeri:16}. Brain surfaces can be modeled as triangulated meshes leading to a more efficient and compact representation. Most existing approaches have analyzed the surfaces of subcortical structures, e.g. hippocampus \cite{Gerardin:09,Ferrarini:09,Costafreda:11} and measure a shape coarsely \cite{wachinger:15}, here we focus on the gray and white matter surfaces.

Recently, spectral signatures based on the eigendecomposition of Laplace-Beltrami Operator (LBO) have attracted a lot of attention in a wide range of applications including object retrieval and classification \cite{Chunyuan:13b,Masoumi:16}. Spectral shape descriptors are grouped into two categories, i.e. global and local descriptors. The local descriptors \cite{Masoumi:17} are defined on each vertex and contain information about the local structure of that vertex, while global descriptors \cite{Reuter:06,Sun:09,Aubry:11} capture information of the entire object. Shape-DNA~\cite{Reuter:06} introduced by Reuter \textit{et al.} as a global signature defined by non-trivial $k$-smallest eigenvalues of LBO normalized by mesh area that sorted in ascending order. Most recently, the authors extended the idea for brain identification called BrainPrint \cite{wachinger:15}. However, a global descriptor only provides a limited representation of a surface as a whole and cannot be applied when, for instance, to analyze local features on surfaces.

Spectral graph wavelet signature (SGWS) is a compact descriptor that is not only pose-invariant but also allows assessing shapes in different scales. In a bid to resolve the deficiencies of the global descriptors, SGWS has been introduced in \cite{Chunyuan:13b,Masoumi:16} as a multiresolution descriptor that encodes both local and global structure of the shape. It efficiently captures valuable information about both macroscopic and microscopic structures of the shape, notably, the cortical folding pattern. SGWS has since showed superiority over ShapeDNA in both non-rigid shape classification and retrieval \cite{Chunyuan:13b,Masoumi:17}. Unlike BrainPrint, which relies on 44 descriptors calculated from both cortical and subcortical surfaces, we compute our SGWS only on the cortical surfaces, with 4 descriptors. This compact representation is less prone to segmentation error and more computationally feasible for large datasets. The new framework leads to more accurate classification and prediction results.

The contributions of this paper are twofold: (1) we propose a framework to precisely model a brain by harnessing the power of the spectral graph wavelets called WaveletBrain. (2) We exploit the WaveletBrain to predict Alzheimer's disease and show the superiority of our approach with respect to the well-known ShapeDNA in different experiments.

\section{Background} \label{Background}
We model a brain surface as a triangular mesh $\mathbb{M}$, where $\mathbb{M}$ is defined by $\mathbb{G}=(\mathcal{V},\mathcal{E},\mathcal{T})$, with $\mathcal{V}=\{\bm{v}_{1},\ldots,\bm{v}_{m}\}$ as the set of vertices, $\mathcal{E}=\{e_{ij}\}$ as the set of edges, and $\mathcal{T}=\{\bm{t}_{1},\ldots,\bm{t}_{g}\}$ as the set of triangles. Each $e_{ij}=[\bm{v}_{i},\bm{v}_{j}]$ connects a pair of vertices $\{\bm{v}_{i},\bm{v}_{j}\}$. We denote two adjacent vertices by $\bm{v}_{i}\sim\bm{v}_{j}$ or simply $i\sim j$) if there is a connectivity between them by an edge, i.e. $e_{ij}\in {\cal E}$. 

\medskip
\noindent{\textbf{Spectral Analysis}}\quad We build our Laplacian matrix by discretization of the Laplace-Beltrami operator (LBO) \cite{Levy:06} based on cotangent weight scheme as suggested by \cite{Meyer:03} given by $\bm{L}=\bm{A}^{-1}(\bm{D-W})$, where $\bm{A}=\mathrm{diag}(a_{i})$ is a mass matrix, $\bm{D}=\mathrm{diag}(d_{i})$ is a degree matrix constructed by $d_{i}=\sum_{j=1}^{n}w_{ij}$, and $\bm{W}=(w_{ij})=\left(\cot\alpha_{ij} + \cot\beta_{ij}\right)/2a_{i}$ is a weight matrix if $i\sim j$ (readers are referred to \cite{Levy:06,Meyer:03} for detailed description).

We solve the \textit{generalized eigenvalue problem}, such that $\bm{C}\bg{\xi}_{\ell}=\lambda_{\ell}\bm{A}\bg{\xi}_{\ell}$, where $\lambda_{\ell}$ and $\bg{\xi}_{\ell}$ are the eigensystem of LBO, and $\bm{C}=\bm{D-W}$.

\section{Method}\label{Method}

\noindent{\textbf{Manifold Harmonic Transform}}\quad For a given graph signal $f: \mathcal{V}\to\mathbb{R}^{m}$, we obtain the eigensystem \{$\lambda_{\ell},\bm{\xi}_{\ell}$\}, $\ell=1,\dots,m$ of LBO to define the manifold harmonic (forward graph Fourier) and inverse manifold harmonic (inverse graph Fourier) transforms as \cite{Hammond:11}

\begin{equation}
\hat{f}(\ell)=\langle f,\xi_{\ell}\rangle=\sum_{i=1}^{m} f(i)\xi_{\ell}(i),\quad 
\text{and}
\quad
f(i)=\sum_{\ell=1}^{m} \hat{f}(\ell)\xi_{\ell}(i),
\label{eq:MHT-IMHT}
\end{equation}
respectively, where $\hat{f}(\ell)$ is the value of $f$ at eigenvalue $\lambda_\ell$ (i.e. $\hat{f}(\ell)=\hat{f}(\lambda_\ell)$).

The lower-order eigenvectors capture the global structure of the brain, while the higher-order eigenvectors encode more details of the surface including cortical folding.

\medskip
\noindent{\textbf{Spectral Graph Wavelet Transform}}\quad Similar to the Fourier transform, the wavelet transform has the power to decompose a signal into its constituent frequencies. However, the advantage of the wavelet transform over the Fourier transform is its capability to perform localization in both frequency and space domain making spectral graph wavelet a perfect candidate for analyzing the signal in multilevel of descriptions.

The localization of wavelet function around a surface point $j\in\mathcal{V}$ may be characterized by applying a wavelet operator $T_{g}^{t}$ at kernel $g$ and scale $t$ to an indicator function $f(i)=\delta_j(i)$ such that $\bg{\psi}_{t,j}=T_{g}^{t}\delta_j$. The inner product between the input function $f$ and $\bg{\psi}_{t,j}$ results in wavelet coefficients as

\begin{equation}
W_{\delta_j}(t,j)=\langle \bg{\delta}_{j},\bg{\psi}_{t,j} \rangle=\sum_{\ell=1}^{m}g(t\lambda_\ell)\xi_{\ell}^{2}(j),
\label{DeltaW_coefficients}
\end{equation}
also we may construct the coefficients of the scaling function by applying the scaling operator $T_{h}$ on a unit impulse function $\delta_j$, i.e. $\bg{\phi}_{j}=T_{h}\delta_j$ as follows 

\begin{equation}
S_{\delta_j}(j)=
\langle \bg{\delta}_{j},\bg{\phi}_{t} \rangle=
\sum_{\ell=1}^{m}h(\lambda_\ell)\xi_{\ell}^{2}(j).
 \label{DeltaS_coefficients}
\end{equation}
We integrate the coefficients of the wavelet and scaling function to build the spectral graph signature at vertex index $j$ as follows:

\begin{equation}
\bm{s}_{L}(j)=\{W_{\delta_j}(t_k,j)\mid k=1,\dots,L\}\cup\{S_{\delta_j}(j)\}.
 \label{Eq:SGWSignatureLevel}
\end{equation}

where $\bm{s}_{L}(j)$ is the spectral graph wavelet descriptor at resolution level $L$. 
The dimension $p$ of SGWS is $L+1$, which leads to a compact signature for analyzing the cortex surface. 
Moreover, to access the full spectrum of the brain structure, the wavelet scales $t_k$ ($t_k > t_{k+1}$) are selected to be logarithmically equispaced between maximum and minimum scales $t_1$ and $t_L$, respectively. 

We define the spectral graph wavelet generating kernel $g$ and scaling function $h$ as follows $g(x)=\frac{2}{\sqrt{3}\pi^\frac{1}{4}} \left(1-x^{2} \right) \exp \left(\frac{-x^{2}}{2} \right)$ and $h(x)=\gamma \exp\left(-(\frac{x}{ 0.3\lambda_{\min}})^4\right)$, respectively.

\begin{figure}[t]
\setlength{\tabcolsep}{.3em}
\centering
\begin{tabular}{ccccc}
\includegraphics[scale=.24]{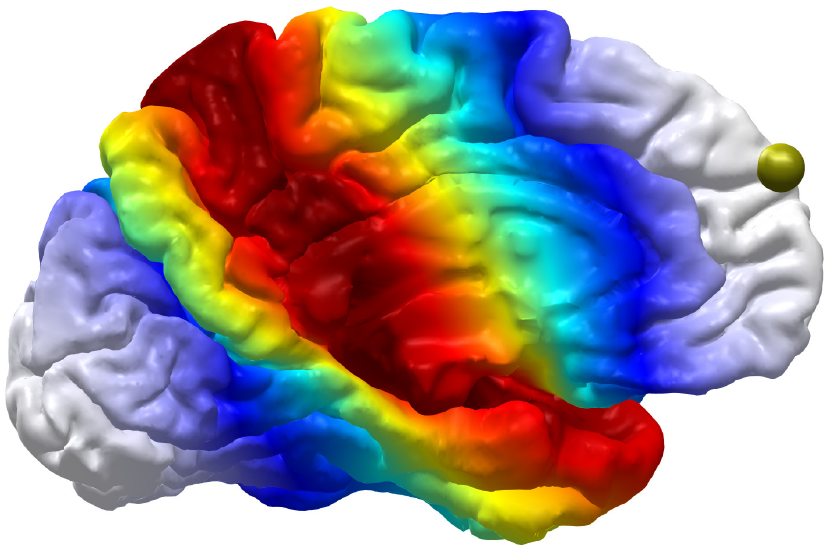}&
\includegraphics[scale=.24]{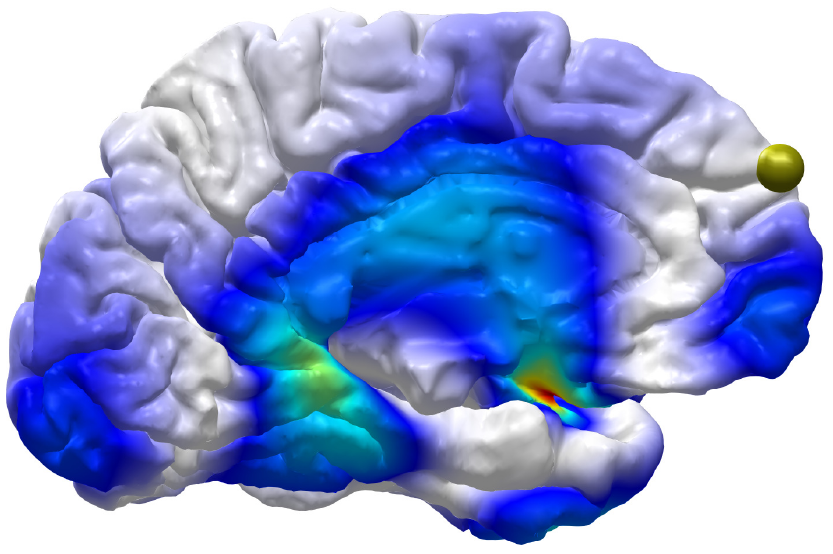}&
\includegraphics[scale=.24]{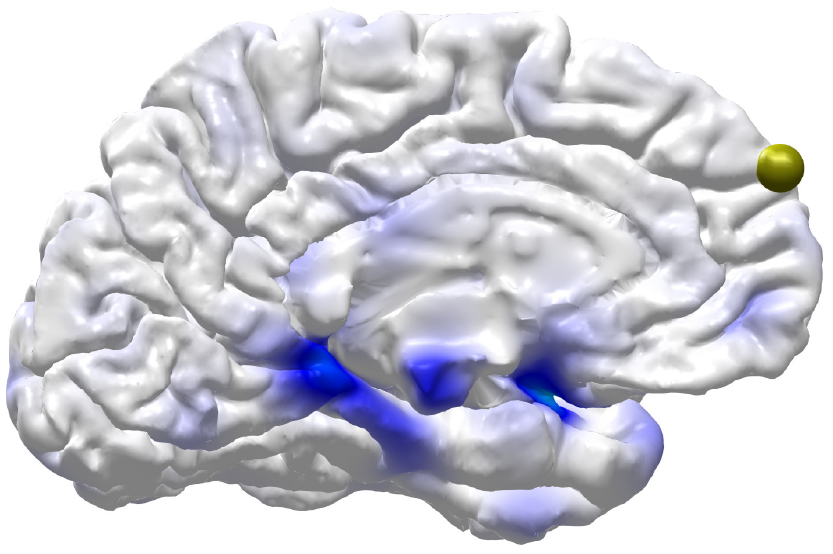}&
\includegraphics[scale=.24]{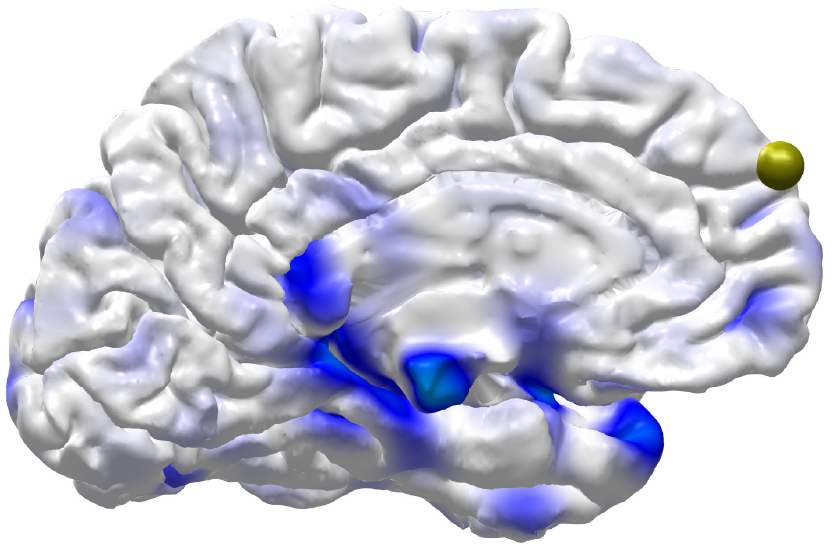}& \multirow{2}{*}{\includegraphics[height=2.1cm]{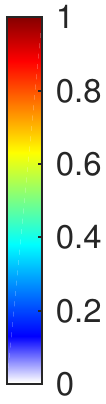}}\\
\includegraphics[scale=.24]{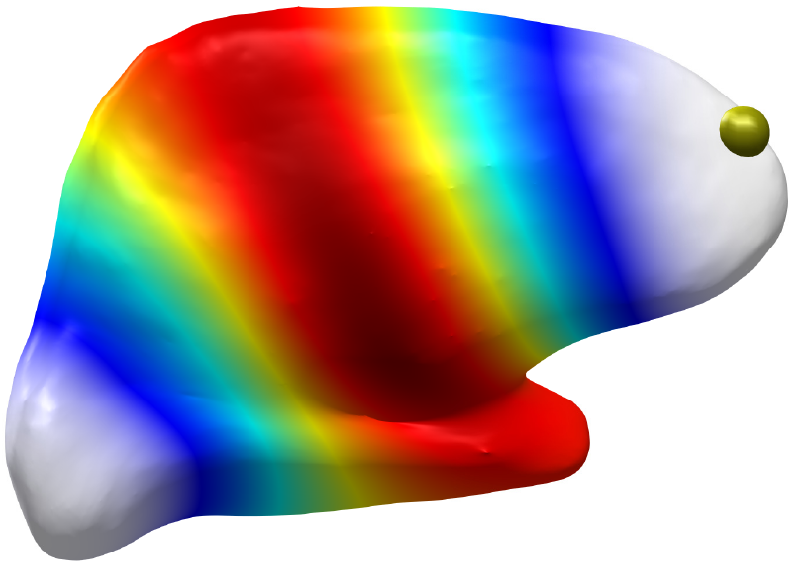}&
\includegraphics[scale=.24]{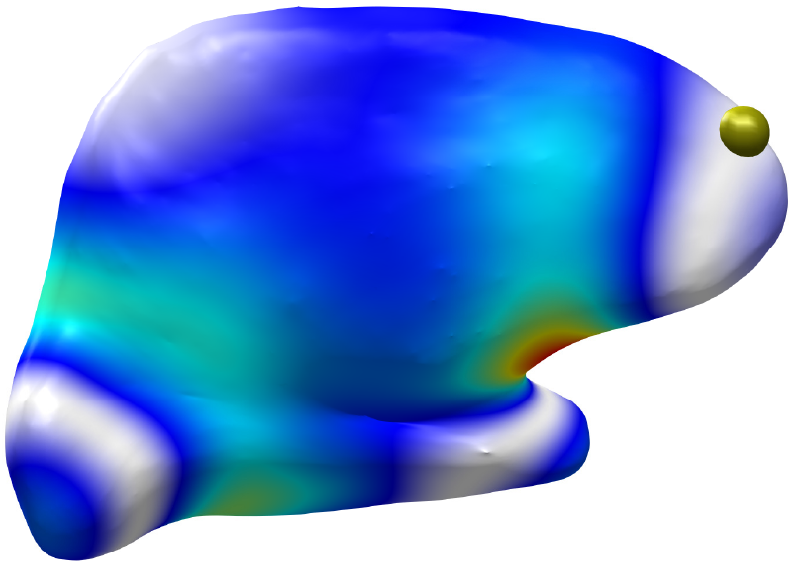}&
\includegraphics[scale=.24]{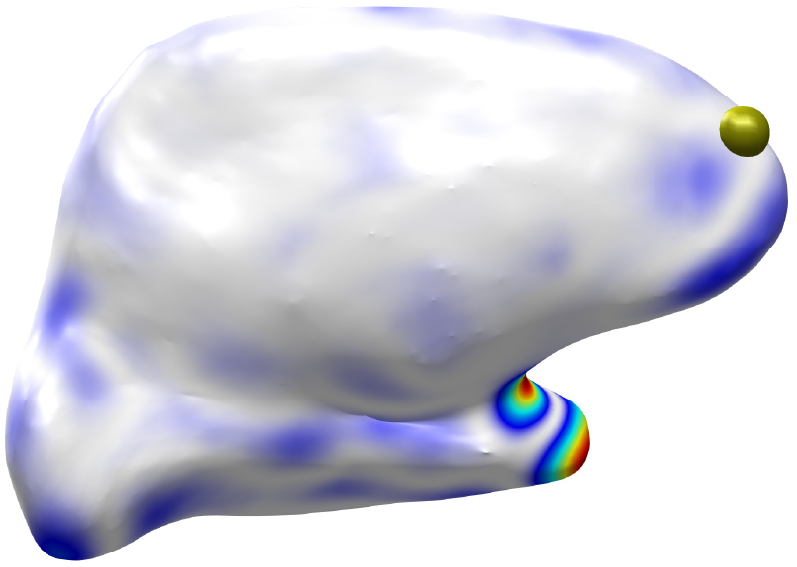}&
\includegraphics[scale=.24]{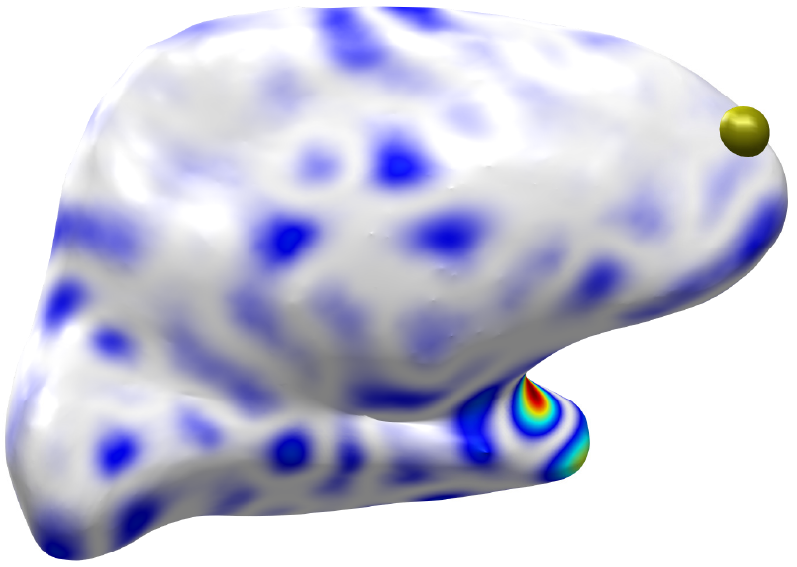}& \\
(a) & (b) & (c) & (d) & 
\end{tabular}
\caption{Representation of the local spectral graph wavelets from a reference point on left pial (top row) and inflated surfaces (bottom row) at $L=3$. Information captured by (a) scaling function, (b)-(d) wavelets, scales 1-3. The colder colors represent a similar structure to the highlighted point while the hotter colors capture the different structure of the brain. Spectral graph wavelets have been visualized on the inflated pial surface for a clearer representation of their variations.}
\label{SGW_representation}
\end{figure}

We set $\gamma$ such that $h(0)$ has the same value as the maximum value of $g$ and $\lambda_{\min}=\lambda_{\max}/15$. Also, the maximum and minimum scales are set to $t_{1}=2/\lambda_{\min}$ and $t_{L}=2/\lambda_{\max}$, where $\lambda_{\min}$ and $\lambda_{\max}$ are the lower and upper bounds of the spectrum, respectively. Figure \ref{SGW_representation} represents the normalized $\chi$-squared distance between a specified point and the rest of points on normal (top row) and smoothed (bottom row) brain surface using SGWS.

\medskip
\noindent{\textbf{Feature Aggregation}}\quad In WaveletBrain, a brain surface $\mathbb{M}$ is modeled by a $p \times m$ matrix $\bm{S}=(\bm{s}_1,\dots,\bm{s}_m)$ of spectral graph wavelet signatures, where each $p$-dimensional feature vector $\bm{s}_{i}$ is a local descriptor that encodes the local information around the $i$th vertex. To aggregate the local features attained from spectral graph wavelet transform (SGWT), we employ bag-of-features (BoF) model leading to a simple and lightweight representation of a brain surface that facilitates the process of brain comparison. The BoF model consists of four main steps: feature extraction and description, dictionary design, feature coding, and pooling.

\begin{figure}[t]
\centering
\includegraphics[scale=0.55]{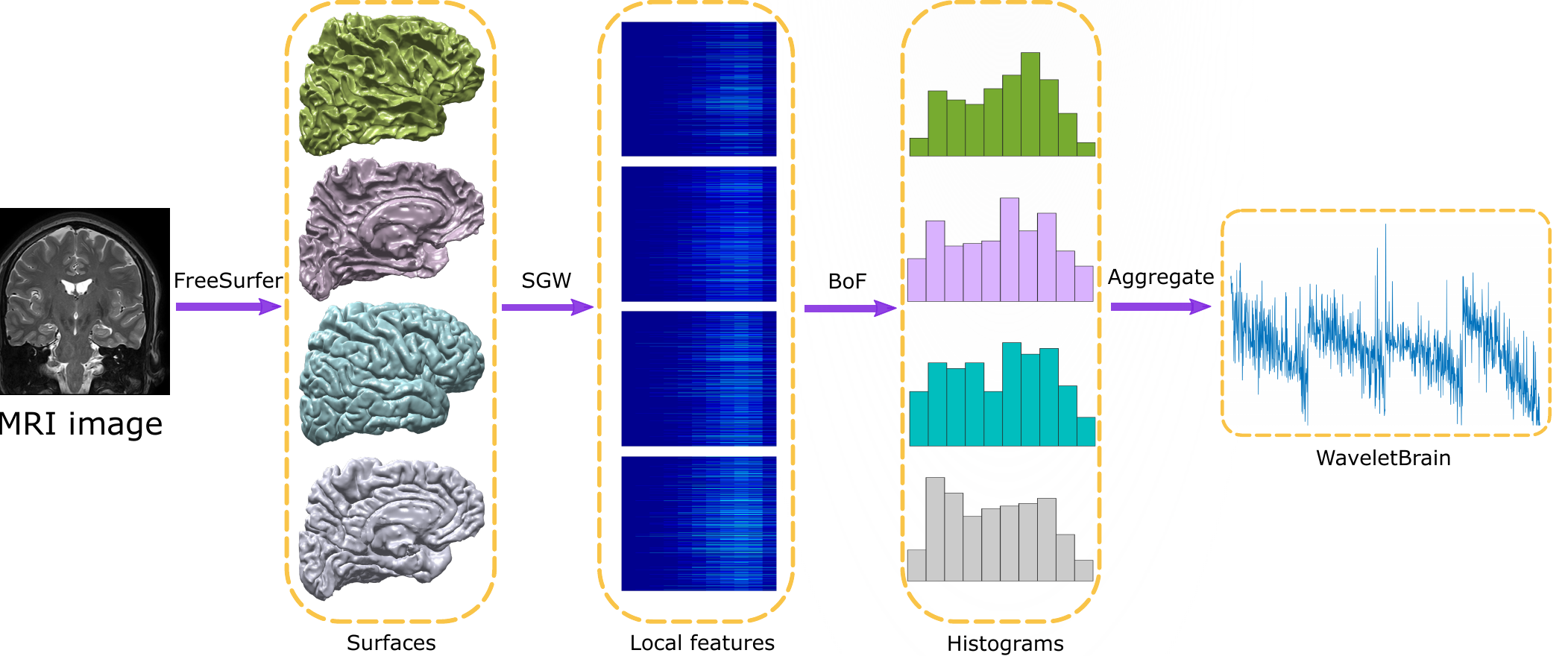}
\caption{Flowchart of the proposed approach. First, brain surfaces of the left/right gray and white matters are created by FreeSurfer. Second, SGW signatures are computed on the surfaces. Third, a global signature (histogram) is calculated by the BoF paradigm for each surface. Fourth, histograms are aggregated to constitute the \it{WaveletBrain}.}
\label{flowchart}
\end{figure}

\medskip
\noindent{\textbf{Algorithm}}\quad In an effort to model the brain shape in spectral graph wavelets domain, we first select the white and gray matter surfaces for each subject in the dataset $\mathcal{D}$. Afterward, each brain surface is characterized to a compact and discriminative representation with SGWS. Subsequently, we perform soft-assignment coding by embedding the local signatures into the geometric dictionary space of size $k$, resulting in higher dimensional mid-level features $\bm{U}=(\bm{u}_{1},\dots,\bm{u}_{m})\in\mathbb{R}^{k\times m}$ which we refer to as spectral graph wavelet codes (SGWC). Consequently, we apply sum-pooling on code assignment matrix $\bm{U}$ to achieve a histogram $h_{r}=\sum_{i=1}^{m}u_{ri}$ that represents each brain surface $\mathbb{M}$. This process is repeated separately for right/left gray and white matters of a brain surface in the dataset. We combine extracted signatures of $n$ brain surfaces in the dataset and arranged them into a $4k\times n$ data matrix $\bm{Z}=(\bm{z}_{1},\dots,\bm{z}_{n})$ called WaveletBrain. It is noteworthy to mention that the dictionary is computed offline by concatenating all the SGWS matrices into a data matrix, followed by applying the K-means algorithm resulting in a vocabulary matrix $\bm{V}$, where $\bm{V}=(\bm{v}_{1},\dots,\bm{v}_{k})\in\mathbb{R}^{p\times k}$. Any further processing such as age prediction and disease classification is conducted on the lightweight representation, i.e. the WaveletBrain $\bm{Z}$. This requires considerably less memory and makes it easier to model, compute and compare than the image scans. WaveletBrain represents shape information in multiple levels of details and offers a greater descriptive representation on, for instance, fine levels of details such as the cortical folds, where definitely important shape information on brain atrophy may be hidden. To provide more discriminative power for WaveletBrain, we normalized our descriptor with surface area $A$, where $\bm{A}=\mathrm{diag}(a_{i})$. The flowchart of the proposed framework is depicted in Figure~\ref{flowchart}.

\section{Experiments and Results}\label{experiment}
In this section, we assess the performance of the proposed  WaveletBrain through extensive experiments for disease classification, sex classification and age prediction. We validate  our approach by performing a comprehensive comparison with ShapeDNA \cite{Reuter:06} as a well-known approach in medical shape analysis. 

\medskip
\noindent{\textbf{Dataset}}\quad The performance of the proposed framework is evaluated on ADNI database (\url{adni.loni.usc.edu}). The ADNI was launched in 2003 as a public-private partnership, led by Principal Investigator Michael W. Weiner, MD. The primary goal of ADNI has been to test whether serial magnetic resonance imaging (MRI), positron emission tomography (PET), other biological markers, and clinical and neuropsychological assessment can be combined to measure the progression of mild cognitive impairment (MCI) and early Alzheimer’s disease (AD). Early diagnosis of AD by sensitive and specific markers is a breakthrough to assist clinicians to develop new preventive treatment as well as reducing time and cost of clinical trial.In our experiment, we selected $719$ subjects from ADNI dataset resulting in $2,892$ surfaces including right/left white and gray matters. 

\medskip
\noindent{\textbf{Performance Evaluation Measures}}\quad 
In a bid to perform AD and sex classification, we split our WaveletBrain data matrix $\mathcal{Z}=\{(\bm{z}_{i},y_{i})\}$, where $\bm{z}_{i} \in \mathbb{R}^{4k}$ in two disjoint subsets of training set $\mathcal{Z}_{\textrm{train}}$ for learning, and the test set $\mathcal{Z}_{\textrm{test}}$ for testing. We evaluate the performance of a classifier on test data by building a contingency table. We then extract the accuracy results by dividing the total number of correctly classified examples by the total number of examples. We use support vector machine (SVM) as a classifier in our experiments to separate different categories since it works well in high-dimensional feature space.

To perform age prediction, we employ partial least-squares (PLS) regression \cite{Wold:84}. PLS is beneficial when the number of variables in a data matrix $\bm{Z}$ is significantly larger than the number of observations. The intuition behind employing PLS regression is the failure of multiple linear regression due to multicollinearity among $\bm{Z}$ variables. As a result, the regression is performed on the latent variables. Intuitively, if there is a disparity between the predicted and the real age, it means a patient is suspected of suffering from AD. 

\begin{figure}[t]
\setlength{\tabcolsep}{.3em}
\centering
\begin{tabular}{ccccc}
\includegraphics[scale=.3]{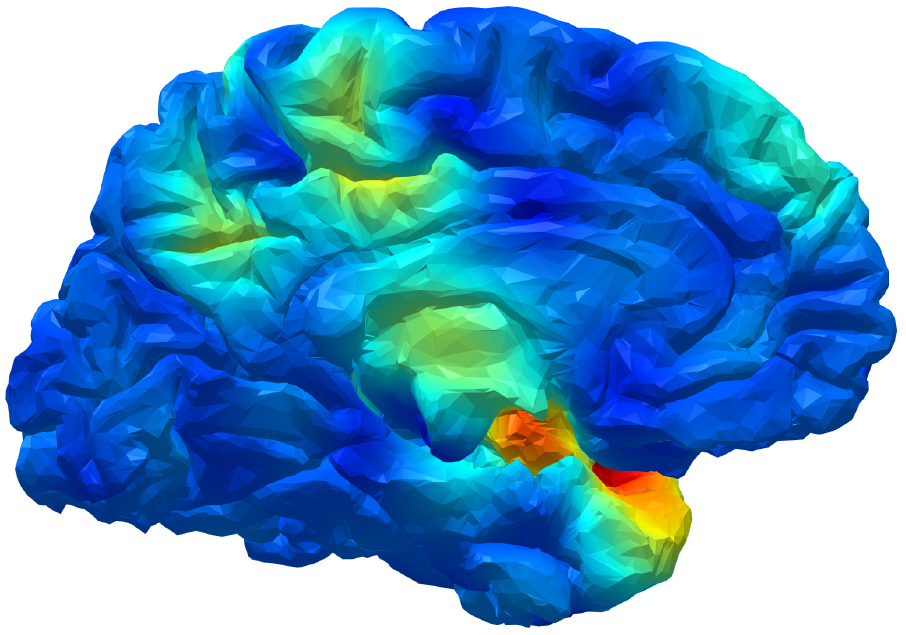}&
\includegraphics[scale=.3]{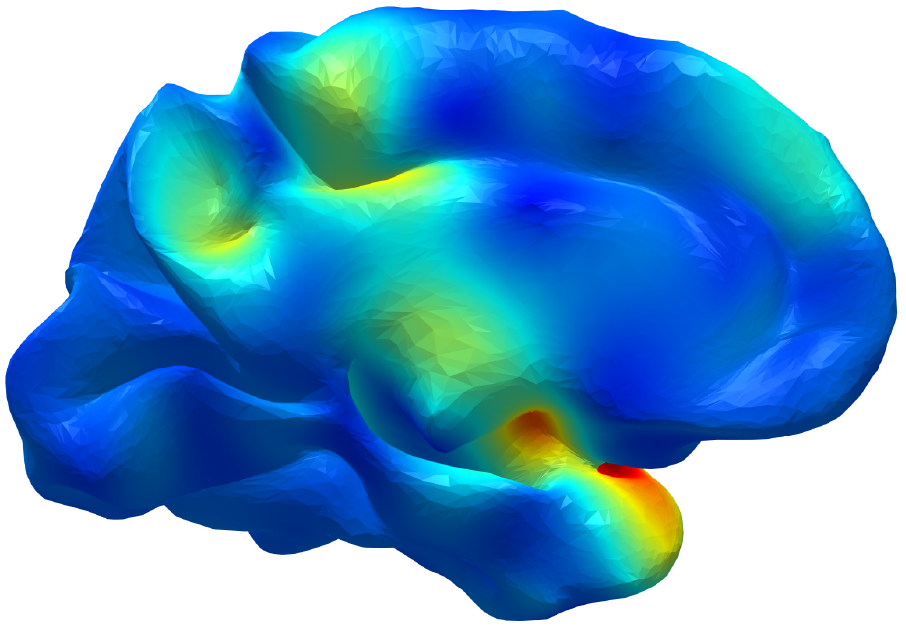}&
\includegraphics[scale=.3]{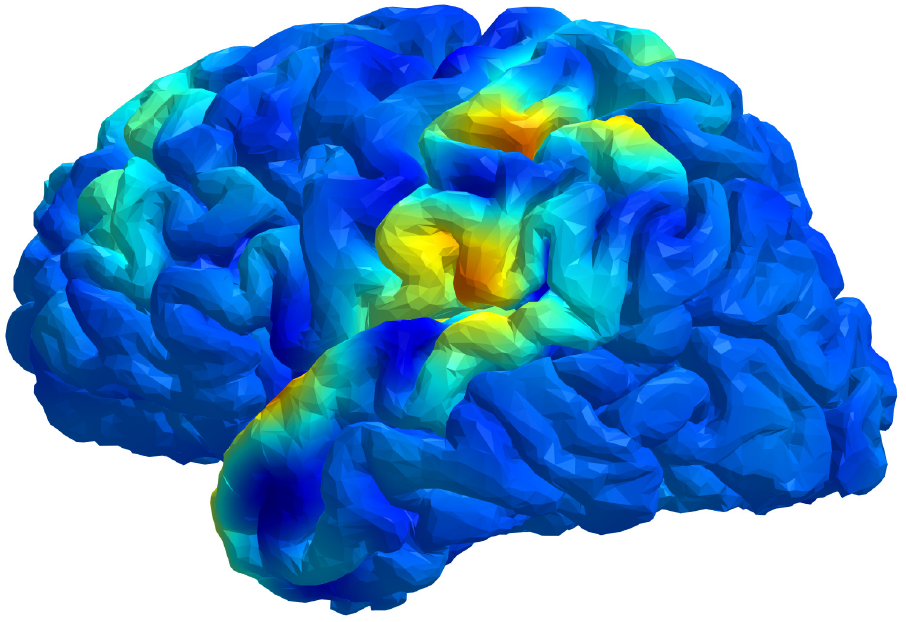}&
\includegraphics[scale=.3]{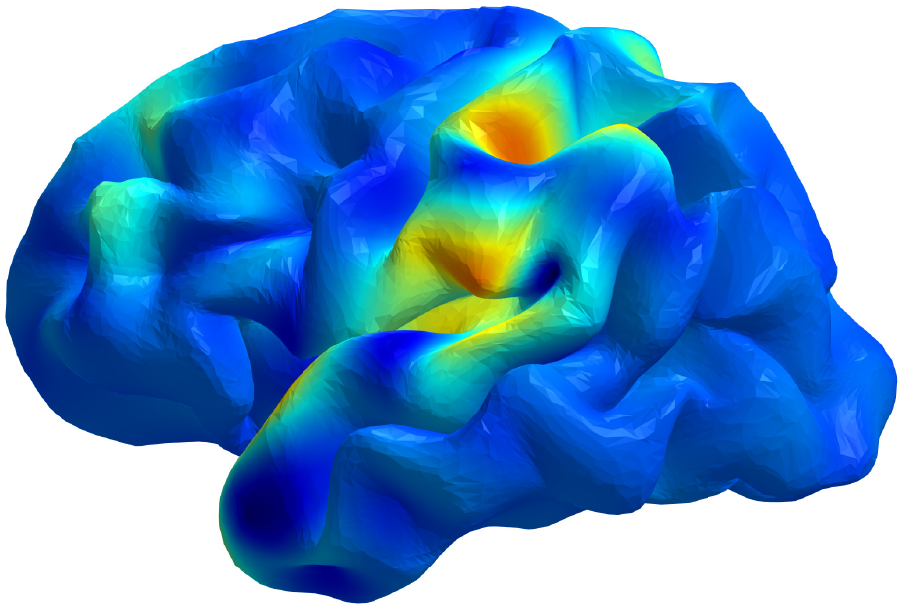}& \multirow{2}{*}{}{\includegraphics[height=2.0cm]{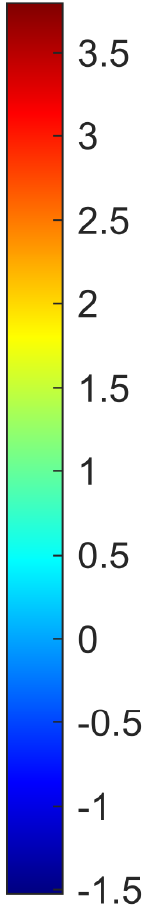}}\\
\multicolumn{2}{c}{medial view} & \multicolumn{2}{c}{lateral view}
\end{tabular}
\caption{Visualization of group differences between AD vs NC on original and smoothed pial surfaces. Coloring indicates differences from low (blue) to high (red).}
\label{Groupdiff_representation}
\end{figure}

\medskip
\noindent{\textbf{Classification of Alzheimer's disease}}\quad
We applied $10$-fold cross-validation to classify brain surfaces of AD from those of the NC and MCI classes in the dataset. We consider an equal size of classes to perform classification. For instance, to classify NC and MCI groups, the data consists of $212$ brain surfaces for each class. To capture more information about the brain shape, we also aggregated the differences between the right gray and white matters as well as those of the left hemispheres to the WaveletBrain. Table \ref{Table:disease classification} reports a comparison between the performance of our WaveletBrain approach with ShapeDNA for differentiating between patients with AD, NC, and MCI. As can be seen, our framework outperforms the ShapeDNA in disease classification and results in higher accuracy with $p$-values of $0.0022$, $0.0001$, and $0.0024$ for AD/NC, NC/MCI, and AD/MCI classification, respectively. For visualization purposes, Fig.~\ref{Groupdiff_representation} reveals the significance map of group differences between AD and NC. As can be seen, the location of brain differences correlates with changes in the hippocampal region.

\begin{table}[b]
\caption{The average accuracy of disease classification for ShapeDNA and WaveletBrain.}
  \label{Table:disease classification}
  \centering
  \resizebox{8cm}{!}{
\begin{tabular}{lllll}
\toprule
 & \quad \# of subjects &\quad ShapeDNA & \quad WaveletBrain & \quad $p$-value\\
 \cmidrule(r){2-5}
AD vs. NC  & \quad $158$ & \quad $65.90 \pm 9.2$  & \quad $\bm{68.75 \pm 7.1}$ & \quad $0.0022$\\
 \midrule
NC vs. MCI & \quad $212$ & \quad $54.49 \pm 6.5$ & \quad $\bm{58.52 \pm 7.3}$ & \quad $0.0001$\\
\midrule
AD vs. MCI & \quad $158$ & \quad $56.00 \pm 4.7$ & \quad $\bm{57.97 \pm 6.6}$ & \quad $0.0024$ \\
\bottomrule
\end{tabular}}
\end{table}

\medskip
\noindent{\textbf{Sex prediction}}\quad
The task here is to predict the sex based on the WaveletBrain. We consider a subset of ADNI dataset to have the same number of male and female subjects, resulting in $N=594$ subjects. The performance of WaveletBrain in sex classification is evaluated by using an SVM classifier. In this experiment, ShapeDNA provided an accuracy of $76.67 \pm 4.70\%$. WaveletBrain achieved an accuracy of $81.67 \pm 4.30\%$, yielding therefore a higher accuracy for sex prediction with a $p$-value of $0.0001$. 

\begin{figure}[b!]
\setlength{\tabcolsep}{.3em}
\centering
\begin{tabular}{cc}
\includegraphics[scale=.35]{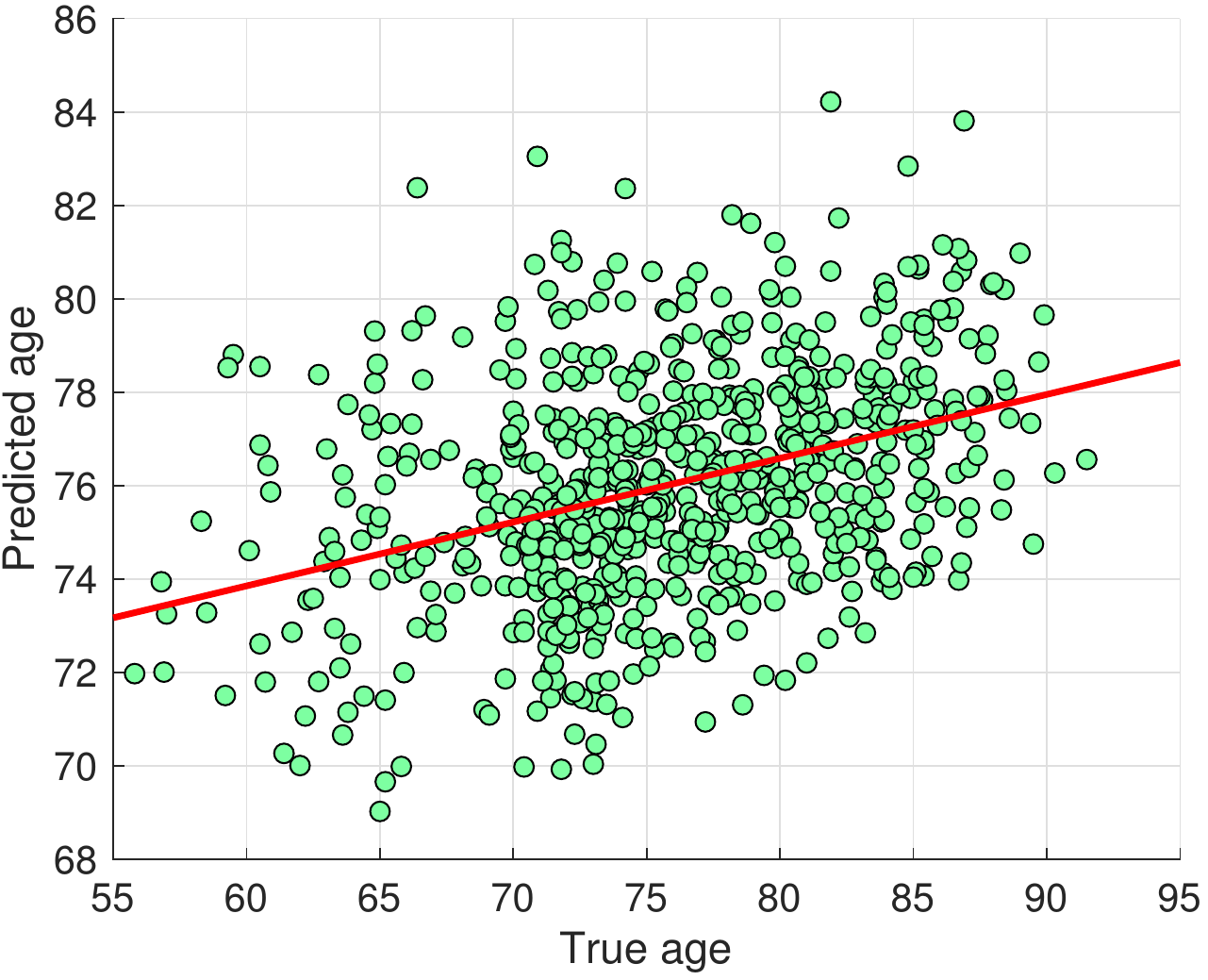}&
\includegraphics[scale=.35]{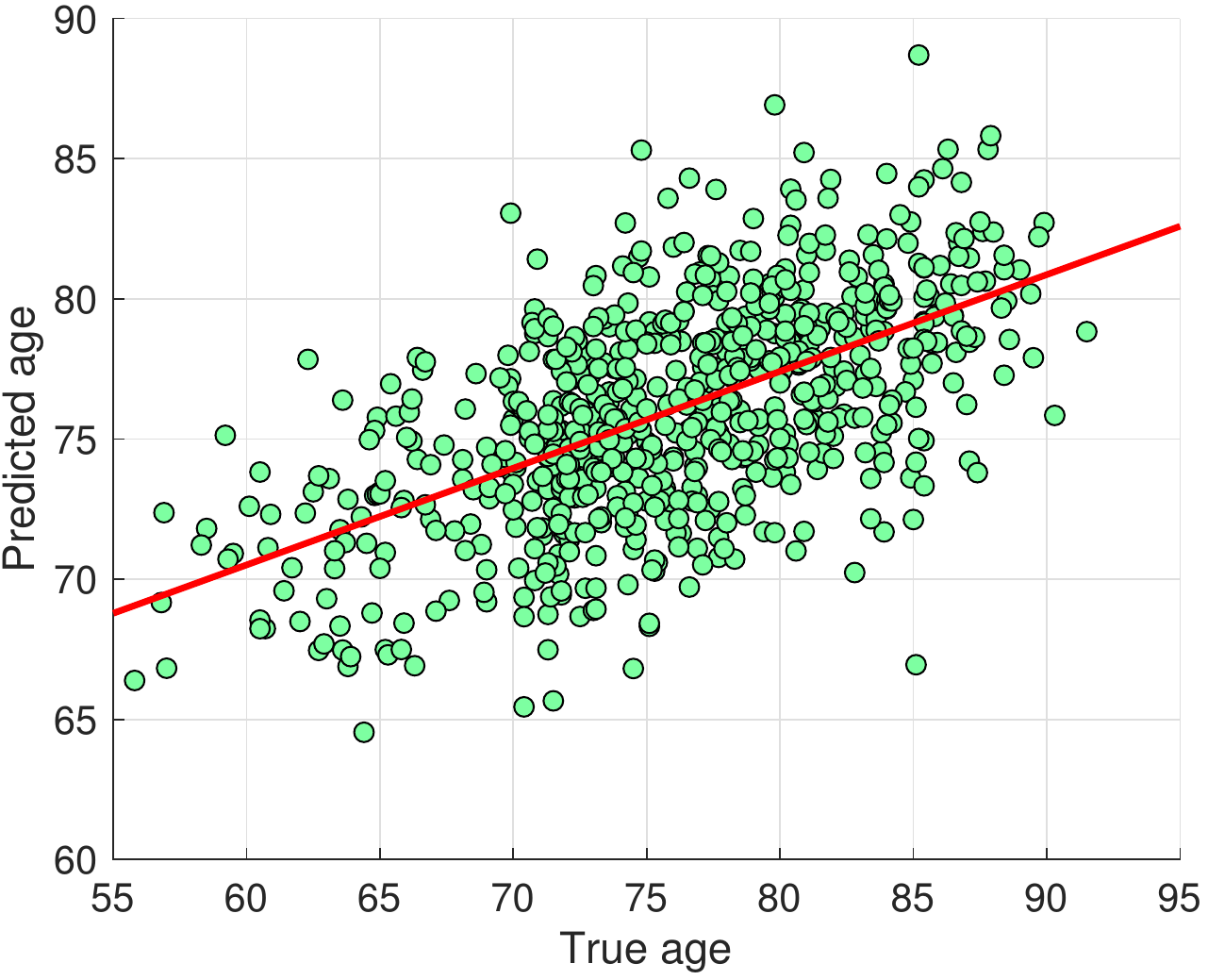}
\end{tabular}
\caption{Scatter plots of true versus predicted ages of subjects in ADNI dataset show the superiority of WaveletBrain (right) with Pearson's $r=0.58$ compared with ShapeDNA (left) with Pearson's $r=0.48$.}
\label{age_prediction}
\end{figure}

\medskip
\noindent{\textbf{Age prediction}}\quad We evaluate the performance of the WaveletBrain on the prediction of the subject's age by exploiting $719$ subjects from ADNI dataset. We employed PLS regression among different diagnostic classes to predict subjects' age. 
Our results are evaluated using a leave-one-out cross-validation. 
Figure \ref{age_prediction} shows a scatter plot of true versus predicted age on the ADNI dataset. The least regression line is also included in the plot for both ShapeDNA and WaveletBrain.

To show the strength of the WaveletBrain in contrast with ShapeDNA for predicting age, we computed Pearson's $r$ for both methods, yielding in $r =0.48$ for ShapeDNA and $r =0.58$ for WaveletBrain, respectively. As can be seen in Figure \ref{age_prediction}, our framework outperforms ShapeDNA in terms of Pearson's $r$ with performance improvement of $0.10$.

In an effort to perform a further comparison between ShapeDNA and WaveletBrain, we calculated the mean absolute error (MAE) for real and estimated ages. The results showed that our proposed method attains an MAE of $4.31$ compared to ShapeDNA with $4.59$. As a result, WaveletBrain ameliorates prediction of the subject's age with a lower error of $0.22$.

\section{Conclusions}
In this paper, we introduced WaveletBrain to characterize brain morphology. We built the WaveletBrain only on the cortical surfaces of the brain resulting in a compact representation that is less prone to segmentation error and ideal for analysis of large datasets. The performance result of the WaveletBrain was compared with the ShapeDNA, yielding in significantly higher accuracy in all experiments including classification of Alz-heimer's disease ($p$-values $\le$ $0.0024$), sex prediction ($p$-values $\le$ $0.0001$) and age prediction (lower MAE error of $0.22$). For future work, we plan to apply the proposed approach to investigate brain asymmetry.
 \section{Acknowledgements}
This work was supported by Fonds de recherche du Quebec Nature et technologies (FQRNT), NSERC and ETS. Data collection and sharing for this project was funded by the Alzheimer’s Disease Neuroimaging Initiative (ADNI) (National Institutes of Health Grant U01 AG024904) and DOD ADNI (Department of Defense award number W81XWH-12-2-0012). ADNI is funded by the National Institute on Aging, the National Institute of Biomedical Imaging and Bioengineering, and through generous contributions from many other sources. Detailed Acknowledgements information is available in \url{http://adni.loni.usc.edu/wp-content/uploads/how_to_apply/ADNI_Manuscript_Citations.pdf}.

%
%
 \bibliographystyle{splncs04}
 \bibliography{biblio}


%
%
%
%
%
\end{document}